\documentclass[11pt,twoside]{article}
\usepackage{mathrsfs}
\usepackage{amsmath}
\usepackage{amssymb}
\usepackage{graphicx}
\usepackage{subfigure}
\usepackage{appendix}
\usepackage{cite}
\numberwithin{equation}{section}

\newtheorem{proposition}{Proposition}

\newtheorem{remark}{Remark}

\DeclareMathOperator{\im}{Im}
\DeclareMathOperator{\re}{Re}
\DeclareMathOperator{\tr}{tr}
\DeclareMathOperator{\diag}{diag}
\newcommand*{\QEDB}{\hfill\ensuremath{\square}}

\title{Integrable boundary conditions for the nonlinear Schr\"{o}dinger hierarchy}
\author{
Baoqiang Xia
\\
School of Mathematics and Statistics, Jiangsu Normal University,
\\
Xuzhou, Jiangsu 221116, P. R. China,
\\
E-mail address: xiabaoqiang@126.com
}

\date{}
\begin{document}
\maketitle
\begin{abstract}

We study integrable boundary conditions associated with the whole hierarchy of nonlinear Schr\"{o}dinger (NLS) equations defined on the half-line. We find that the even order NLS equations and the odd order NLS equations admit rather different integrable boundary conditions. In particular, the odd order NLS equations permit a new class of integrable boundary conditions that involves the time reversal. We prove the integrability of the NLS hierarchy in the presence of our new boundary conditions in the sense that the models possess infinitely many integrals of the motion in involution. Moreover, we develop further the boundary dressing technique to construct soliton solutions for our new boundary value problems.


%\noindent{\bf PACS numbers:}\quad 02.30.Ik, 02.30.Jr
\end{abstract}
\newpage

\section{ Introduction}

The study on characterising integrable boundary conditions for integrable nonlinear PDEs is one of the most important issues in integrable systems. From the point of view of Hamiltonian theory, Sklyanin presented a systematic approach \cite{Sklyanin1987} to identify boundary conditions that preserve the integrability of a model with the space variable posed on the half-line or the finite interval. In Sklyanin's approach, a boundary condition is encoded into a matrix, called boundary $K(\lambda)$ matrix, that satisfies certain algebraic constraint involving the classical $r$-matrix. The integrability of the resulting boundary systems is ensured by the existence of infinitely many integrals of the motion in involution.
Several solution methods to the integrable boundary value problems on the half-line (or the interval) have been developed, such as the so-called nonlinear mirror image method \cite{Habibullin1991,BT1989,BT1991,T1991,BH2009,BB2012,CZ2012,CCD2021}, and a boundary dressing technique presented recently in \cite{Zhang2019,ZZ2021,WZ2022}.
It is worth reminding that the Fokas' unified transform method \cite{Fokas}, as a general method to deal with boundary value problems, can also be used to tackle the problem in the presence of integrable boundary conditions \cite{BFS2015}.

The nonlinear Schr\"{o}dinger (NLS) equation is a mathematically and physically important integrable model. The integrable boundary conditions for the NLS equation have been extensively investigated. There are mainly two different kinds of integrable boundary conditions known for the NLS equation. The first one is the well-addressed Robin boundary condition (see e.g. \cite{Sklyanin1987}), which includes the Dirichlet and Neumann boundary conditions as limiting cases. The second one is a new boundary condition involving the time derivative of the NLS field that is investigated recently in \cite{Zambon2014,Xia2021,CCD2021,CCRZ2022,Gruner2020,Zhangc2021}.
In aspect of integrability, the main difference between the two boundary conditions is that the first one is characterised by the constant $K(\lambda)$ matrix \cite{Sklyanin1987}, while the second one is characterised by the dynamical $K(\lambda)$ matrix \cite{Zambon2014}.

In this paper, we study integrable boundary conditions for the NLS hierarchy defined on the half-line. The NLS hierarchy is the higher order integrable extension of the classical NLS equation \cite{AKNS1,AKNS2}. The classical NLS equation and the complex modified Korteweg-de Vries (mKdV) equation correspond, respectively, to the first and second nontrivial members in this hierarchy. Our boundary conditions are derived by imposing suitable reductions on the B\"{a}cklund transformations (BTs) for the NLS hierarchy. We note that the idea to study boundary conditions compatible with the integrability via BTs was initiated by Habibullin in \cite{Habibullin1991}. Here we generalize this technique to the whole NLS hierarchy. We find that the even order NLS equations and the odd order NLS equations in the hierarchy admit structurally rather different integrable boundary conditions. For the even order NLS equations, we obtain two classes of integrable boundary conditions: one is the higher order generalization of the well-known Robin boundary condition and the other one is the generalization of a new boundary condition received less investigation (see section 3.1). For the odd order NLS equations, we find a new type of integrable boundary conditions that involves the time reversal (see section 3.2). To the best of our knowledge, this kind of boundary conditions have not reported in the literature yet. For all our new boundary conditions, we establish the integrability in the sense of the existence of infinitely many integrals of the motion in involution. In particular, the $K(\lambda)$ matrices describing our new boundary conditions are constructed and infinitely many Poisson commuting conserved quantities for the model with our new boundary conditions are derived explicitly by investigating the large $\lambda$ expansions of the trace of the corresponding monodromy matrices.

In this paper, we also show how to construct soliton solutions meeting the new integrable boundary conditions we presented. The method we use is the boundary dressing technique introduced in \cite{Zhang2019,ZZ2021,WZ2022}, which is based on the tool of Darboux transformations (DTs) for integrable systems (see e.g. \cite{MS1991,Gu}). Our results provide an extension of the results in \cite{Zhang2019,ZZ2021,WZ2022} from the case with boundary conditions characterised by constant $K(\lambda)$ matrices to the case with boundary conditions characterised by dynamical $K(\lambda)$ matrices and to the case with boundary conditions involving time reversal.

The paper is arranged as follows. In section 2, we briefly review the hierarchy of NLS equations and the DTs and BTs associated with this hierarchy. In section 3, we construct possible integrable boundary conditions for the whole NLS hierarchy by virtue of the corresponding BTs in conjunction with suitable reductions. In section 4, we establish the integrability of the new boundary conditions presented in section 3 via Sklyanin's approach. In section 5, we develop further the boundary dressing technique to construct soliton solutions of the NLS hierarchy in the presence of our new boundary conditions.

\section{The NLS hierarchy and its Darboux, B\"{a}cklund transformations}

\subsection{The hierarchy of NLS equations}

We consider the following auxiliary linear problems (usually called as Lax pair) \cite{AKNS1,AKNS2}
\begin{subequations}
\begin{eqnarray}
\phi_x(x,t,\lambda)=U(x,t,\lambda)\phi(x,t,\lambda),
~~
U(x,t,\lambda)=\left( \begin{array}{cc} \frac{\lambda}{2i} & -\bar{u} \\
 u &  -\frac{\lambda}{2i} \\ \end{array} \right),
 \label{lpx}
\\
\phi_t(x,t,\lambda)=V^{(n)}(x,t,\lambda)\phi(x,t,\lambda),
~~ V^{(n)}(x,t,\lambda)=\sum_{j=0}^n\left( \begin{array}{cc}  a_j & b_j \\
 c_j & -a_j  \\ \end{array} \right)\lambda^{n-j},
 \label{lpt}
\end{eqnarray}
\label{lpxt}
\end{subequations}
where $\lambda$ is a spectral parameter, and $a_j$, $c_j$ and $b_j=-\bar{c}_j$ are defined recursively by
\begin{eqnarray}
\begin{split}
a_0=\frac{i}{2},~~b_0=c_0=0,
\\
a_1=0,~~b_1=\bar{u},~~c_1=-u,
\\
\left( \begin{array}{c} c_{j+1} \\
  b_{j+1} \\ \end{array} \right)
=\left( \begin{array}{cc} -i\partial_x-2iu\partial_x^{-1}\bar{u} & -2iu\partial_x^{-1}u \\
 2i\bar{u}\partial_x^{-1}\bar{u} &  i\partial_x+2i\bar{u}\partial_x^{-1}u \\ \end{array} \right)
 \left( \begin{array}{c} c_{j} \\
  b_{j} \\ \end{array} \right),~~j\geq 1,
\\
a_{j+1}=-\partial_x^{-1}\left(\bar{u}c_{j+1}+ub_{j+1}\right),~~j\geq 1.
\end{split}
\label{abc}
\end{eqnarray}
In the above expressions we have used the following conventions: the bar indicates complex conjugation, $\partial_x^{-1}$ stands for the inverse of the operator  $\partial_x$, and we take $\partial_x^{-1}0=0$.
Let us list the first few members of $a_j$ and $c_j$ that will be used,
\begin{eqnarray}
\begin{split}
a_2=-i|u|^2,~~c_2=iu_x,
\\
a_3=u\bar{u}_x-u_x\bar{u},~~c_3=u_{xx}+2|u|^2u,
\\
a_4=i\left(u\bar{u}_{xx}+\bar{u}u_{xx}-|u_x|^2+3|u|^4\right),~~c_4=-i\left(u_{xxx}+6|u|^2u_x\right).
\end{split}
\label{abc234}
\end{eqnarray}
We can check directly that the compatibility condition of (\ref{lpx}) and (\ref{lpt}), that is
\begin{eqnarray}
U_t-V^{(n)}_x+[U,V^{(n)}]=0,
\label{zce}
\end{eqnarray}
gives rise to a hierarchy of evolution equations
\begin{eqnarray}
u_t=ic_{n+1}.
\label{nlsh}
\end{eqnarray}
For example, for $n=2$, equation (\ref{nlsh}) becomes the classical NLS equation
\begin{eqnarray}
iu_t+u_{xx}+2 u|u|^2=0.
\label{nls}
\end{eqnarray}
For $n=3$, equation (\ref{nlsh}) becomes the complex mKdV equation %\cite{AKNS1,AKNS2}
\begin{eqnarray}
u_t-u_{xxx}-6 |u|^2u_x=0.
\label{cmkdv}
\end{eqnarray}
For $n=4$, equation (\ref{nlsh}) yields the fourth order NLS equation
\begin{eqnarray}
iu_t-u_{xxxx}-8|u|^2u_{xx}-6\bar{u}u_x^2-4u|u_x|^2-2u^2\bar{u}_{xx}-6 u|u|^4=0.
\label{fnls}
\end{eqnarray}
For the general $n$, we will refer to (\ref{nlsh}) as $n$th NLS equation.

\subsection{Darboux transformations}

The DT can be seen as a gauge transformation that leaves the Lax system invariant. We use $u[0]$ to denote the original potential function and $U[0]$, $V^{(n)}[0]$ and $\phi[0]$ to describe the corresponding undressed Lax system (\ref{lpxt}).
%We use $u[0]$ and $u[1]$ to denote respectively the original potential function and the transformed one, and use $(U[0], V^{(n)}[0], \phi[0])$ and $(U[1], V^{(n)}[1], \phi[1])$ to describe the undressed Lax pair system (\ref{lpxt}) and the transformed one respectively.
Let $\psi_1=\left(\mu_1,\nu_1\right)^T$ be a special solution of the undressed Lax system (\ref{lpxt}) fixed at  $\lambda=\lambda_1$. Construct the matrix
\begin{eqnarray}
D[1](x,t,\lambda)=(\lambda-\bar{\lambda}_1)\mathbf{I}+\left(\bar{\lambda}_1-\lambda_1\right)P[1],
~~P[1]=\frac{\psi_1\psi^\dag_1}{\psi^\dag_1\psi_1},
\label{DTM1}
\end{eqnarray}
where $\mathbf{I}$ stands for the $2\times 2$ identity matrix, and the superscript $\dag$ denotes the Hermitian, i.e. complex conjugate transpose.
The DT for the NLS hierarchy is defined by the following gauge transformation \cite{MS1991,Gu}
\begin{subequations}
\begin{eqnarray}
\phi[1]=D[1](x,t,\lambda)\phi[0],
\label{DTa}
\\
u[1]=u[0]+i\left(\lambda_1-\bar{\lambda}_1\right)\frac{\bar{\mu}_{1}\nu_{1}}{|\mu_1|^2+|\nu_1|^2},
\label{DTb}
\end{eqnarray}
\label{DT}
\end{subequations}
which preserves the forms of the Lax system (\ref{lpxt}).
%In other words, under the action of (\ref{DT}), the transformed system $(U[1], V^{(n)}[1], \phi[1])$ is structurally identical to the undressed system.
Using (\ref{DTb}), it is easy to compute one-soliton solutions for the NLS hierarchy starting from the zero seed solution. For example, the solution $\psi_1$ corresponding to the undressed Lax system for the NLS equation is
\begin{eqnarray}
\psi_1(x,t,\lambda_1)=\left(\mu_1,\nu_1\right)^T=e^{-\frac{i}{2}(\lambda_1x-\lambda^2_1t)\sigma_3}\left(u_1,v_1\right)^T, \label{psinls}
\end{eqnarray}
where $\left(u_1,v_1\right)^T$ is a nonzero constant vector. Inserting (\ref{psinls}) into (\ref{DTb}), we obtain the following on-soliton solution for the NLS equation
\begin{eqnarray}
u=\im(\lambda_1)
\frac{\exp\left\{i\left[\re(\lambda_1)x-\left((\re(\lambda_1))^2-(\im(\lambda_1))^2\right)t+\theta\right]\right\}}
{\cosh\left\{\im(\lambda_1)\left[x-2\re(\lambda_1)t\right]+x_0\right\}},
\label{u1nls}
\end{eqnarray}
where $\exp\{i\theta\}=-\frac{\bar{u}_1v_1}{|u_1||v_1|}$ and $\exp\{x_0\}=\frac{|u_1|}{|v_1|}$. Similarly, the one-soliton solution for the complex mKdV equation resulting from the zero seed solution reads
\begin{eqnarray}
u=\im(\lambda_1)
\frac{\exp\left\{i\left[\re(\lambda_1)\left(x-\left((\re(\lambda_1))^2-3(\im(\lambda_1))^2\right)t\right)+\theta\right]\right\}}
{\cosh\left\{\im(\lambda_1)\left[x-\left(3(\re(\lambda_1))^2-(\im(\lambda_1))^2\right)t\right]+x_0\right\}}.
\label{u1mkdv}
\end{eqnarray}

The DT can be done successively. We write the $N$-times transformed system as $U[N]$, $V^{(n)}[N]$ and $\phi[N]$. Let $\psi_j=\left(\mu_j,\nu_j\right)^T$, $j=1,\cdots,N$, be column solutions of the undressed Lax system (\ref{lpxt}) fixed at $\lambda=\lambda_j$, $j=1,\cdots,N$, respectively. The $N$-step DT matrix takes the form
\begin{eqnarray}
D[N](x,t,\lambda)=\left((\lambda-\bar{\lambda}_N)\mathbf{I}+\left(\bar{\lambda}_N-\lambda_N\right)P[N]\right)\cdots
\left((\lambda-\bar{\lambda}_1)\mathbf{I}+\left(\bar{\lambda}_1-\lambda_1\right)P[1]\right),
\label{DTMN}
\end{eqnarray}
where
\begin{eqnarray}
P[j]=\frac{\psi_j[j-1]\psi^\dag_j[j-1]}{\psi^\dag_j[j-1]\psi_j[j-1]},
~~\psi_j[j-1]=\left.D[j-1]\right|_{\lambda=\lambda_j}\psi_j.
\end{eqnarray}
The newly transformed potential function corresponding to the $N$-step DT is given by
\begin{eqnarray}
u[N]=u[0]-i\left(\Sigma_1\right)_{21},
\label{un}
\end{eqnarray}
where $\left(\Sigma_1\right)_{21}$ stands for the $(21)$-element of the matrix $\Sigma_1$, and $\Sigma_1$ is the matrix coefficient of $\lambda^{N-1}$ of $D[N]$, i.e.
\begin{eqnarray}
\Sigma_1=-\sum_{j=1}^N\left(\bar{\lambda}_j\mathbf{I}-\left(\bar{\lambda}_j-\lambda_j\right)P[j]\right).
\label{Sig1}
\end{eqnarray}
%Formula (\ref{un}) can be expressed in the following form
%\begin{eqnarray}
%u[N]=u[0]-i\frac{\Delta_1}{\Delta_2},
%\label{un2}
%\end{eqnarray}
%where.
The above formulae can be used to compute $N$-soliton solutions starting from a properly chosen seed solution $u[0]$ (such as the zero seed solution); see e.g. \cite{MS1991,Gu} for details.

We note that the DT matrix $D[N](x,t,\lambda)$ satisfies the following symmetry relations
\begin{eqnarray}
\left(D[N](\lambda)\right)_{11}=\overline{\left(D[N](\bar{\lambda})\right)_{22}},
~~\left(D[N](\lambda)\right)_{12}=-\overline{\left(D[N](\bar{\lambda})\right)_{21}},
\label{symdt}
\end{eqnarray}
where $\left(D[N]\right)_{jk}$, $j,k=1,2$, stand for the $jk$-elements of the matrix $D[N]$.

\subsection{B\"{a}cklund transformations}

The DT matrix (\ref{DTM1}) can be expressed in terms of the spectral parameter $\lambda$
and the old and new potentials $u[0]$ and $u[1]$.
Indeed, using (\ref{DTb}) we may express the off-diagonal elements of $D[1]$ as
\begin{eqnarray}
\left(D[1]\right)_{21}=i\left(u[1]-u[0]\right),~~\left(D[1]\right)_{12}=i\left(\overline{u[1]}-\overline{u[0]}\right).
\label{odd1}
\end{eqnarray}
Using the following equalities
\begin{eqnarray}
\left(D[1]\right)_{11}+\left(D[1]\right)_{22}=2\lambda-\left(\lambda_1+\bar{\lambda}_1\right),
~~\det D[1]=\left(\lambda-\lambda_1\right)\left(\lambda-\bar{\lambda}_1\right),
\end{eqnarray}
we find the diagonal elements of $D[1]$ can be expressed as
\begin{eqnarray}
\left(D[1]\right)_{11}=\lambda-\left(\re \lambda_1\pm i\sqrt{(\im\lambda_1)^2 - |u[1]-u[0]|^2}\right),
~\left(D[1]\right)_{22}=2\lambda-\lambda_1-\bar{\lambda}_1-\left(D[1]\right)_{11}.
\label{dd1}
\end{eqnarray}
Without loss of generality, we fix the signs in $\left(D[1]\right)_{11}$ to be positive.
To simplify further the notation, we write $u[0]$ and $u[1]$ as $\tilde{u}$ and $u$ respectively,
and we set the complex parameter $\lambda_1=\alpha+i\beta$, $\alpha,\beta\in\mathbb{R}$.
Then we obtain from (\ref{DTM1}) the matrix
\begin{eqnarray}
\mathcal{D}=\lambda\mathbf{I}+\left( \begin{array}{cc} -\alpha-i\Omega & -i(\bar{\tilde{u}}-\bar{u})\\
 -i (\tilde{u}-u) &  -\alpha+i\Omega  \\ \end{array} \right),
~~\Omega=\sqrt{\beta^2 - |u-\tilde{u}|^2},
 \label{nlsdm}
\end{eqnarray}
with $\alpha$ and $\beta$ being real parameters.

The Darboux matrix $\mathcal{D}$ transforms the auxiliary function as $\phi=\mathcal{D}\tilde{\phi}$.
The invariance of the Lax pair under the transformation yields the equations for the BT:
\begin{subequations}
\begin{eqnarray}
\mathcal{D}_x(x,t,\lambda)=U(x,t,\lambda)\mathcal{D}(x,t,\lambda)-\mathcal{D}(x,t,\lambda)\tilde{U}(x,t,\lambda),
\label{BT1a}
\\
\mathcal{D}_t(x,t,\lambda)=V^{(n)}(x,t,\lambda)\mathcal{D}(x,t,\lambda)-\mathcal{D}(x,t,\lambda)\tilde{V}^{(n)}(x,t,\lambda).
\label{BT1b}
\end{eqnarray}
\label{BT1}
\end{subequations}
Inserting (\ref{nlsdm}) into the above set of equations, we obtain the explicit form of BT for the $n$th NLS equation:
\begin{subequations}
\begin{eqnarray}
\tilde{c}_j-c_j-i\left(\tilde{u}-u\right)\left(\tilde{a}_{j-1}+a_{j-1}\right)
+i\Omega\left(\tilde{c}_{j-1}+c_{j-1}\right)-\alpha\left(\tilde{c}_{j-1}-c_{j-1}\right)=0,
~~2\leq j\leq n,
\label{BTNLSa}
\\
\tilde{u}_t-u_t+\left(\tilde{u}-u\right)\left(\tilde{a}_n+a_n\right)
-\Omega\left(\tilde{c}_{n}+c_{n}\right)-i\alpha\left(\tilde{c}_{n}-c_{n}\right)=0.
\label{BTNLSb}
\end{eqnarray}
\label{BTNLS}
\end{subequations}
In particular, for $n=2$ we recover the following BT (see e.g. \cite{Lamb1974,Chen1974,KW1975})
\begin{eqnarray}
\begin{split}
\tilde{u}_x-u_x=\Omega\left(\tilde{u}+u\right)+i\alpha\left(\tilde{u}-u\right),
\\
\tilde{u}_t-u_t=i\left(\tilde{u}-u\right)\left(|\tilde{u}|^2+|u|^2\right)
+i\Omega\left(\tilde{u}_x+u_x\right)-\alpha\left(\tilde{u}_x-u_x\right),
\end{split}
\label{btnls}
\end{eqnarray}
for the NLS equation (\ref{nls}).

%{\bf Remark} Straightforward computations shows that the BT equations resulting from (\ref{BT1a}) are included in the equations resulting from (\ref{BT1b}). Indeed, . Thus equation (\ref{BT1b}) is sufficient to determine the BT (\ref{BTNLS}).

\section{Boundary conditions arising from reductions of B\"{a}cklund transformations}

In this section we identify possible boundary conditions that are compatible with the integrability of the NLS hierarchy by using the technique of BTs together with suitable reductions. This technique was introduced by Habibullin in \cite{Habibullin1991}. Here the main difference between our method and Habibullin's method is that we consider both the space and time parts of BTs and we consider reductions involving not only space inversion but also time reversal (in Habibullin's method only space parts of the BTs are considered and only space reductions are used). As a consequence, we obtain new integrable boundary conditions for the equations in the NLS hierarchy.

\subsection{Boundary conditions for even order NLS equations}

For $n=2k$, $k\geq 1$, one can check directly that if $u(x,t)$ solves the $n$th NLS equation (\ref{nlsh}),
then so does $\varepsilon u(-x,t)$, $\varepsilon=\pm 1$. This implies that we can consider the following reduction between the two potentials of the $2k$th NLS equation related by BT:
\begin{eqnarray}
\tilde{u}(x,t)=\varepsilon u(-x,t),~~\varepsilon=\pm 1.
\label{red1}
\end{eqnarray}
We will specify the two cases $\varepsilon=1$ and $\varepsilon=-1$ separately, due to they induce two different kinds of boundary conditions.

{\bf Case I. Higher order generalization of the Robin boundary condition corresponding to $\varepsilon=1$.} In the case of $\varepsilon=1$, we may deduce from (\ref{abc}) and (\ref{red1}) that
\begin{eqnarray}
\begin{split}
\left.\tilde{c}_{2j-1}\right|_{x=0}=\left.c_{2j-1}\right|_{x=0},
~~\left.\tilde{c}_{2j}\right|_{x=0}=-\left.c_{2j}\right|_{x=0},
~~1\leq j\leq k,
\\
\left.\tilde{a}_{2j-1}\right|_{x=0}=-\left.a_{2j-1}\right|_{x=0},
~~\left.\tilde{a}_{2j}\right|_{x=0}=\left.a_{2j}\right|_{x=0},
~~1\leq j\leq k.
\end{split}
\label{acred1}
\end{eqnarray}
Using (\ref{acred1}), we find that (\ref{BTNLSb}) evaluated at $x=0$ becomes an identity in the situation $\alpha=0$, while (\ref{BTNLSa}) evaluated at $x=0$ yields the following boundary conditions
\begin{eqnarray}
\left.\left(c_{2j}-i\beta c_{2j-1}\right)\right|_{x=0}=0,~~1\leq j\leq k,
\label{bceh}
\end{eqnarray}
for the $2k$th NLS equation (namely, (\ref{nlsh}) with $n=2k$).
The boundary conditions (\ref{bceh}) can be viewed as a higher order generalization of the Robin boundary condition.
In particular, for $k=1$ we recover the standard Robin boundary condition
\begin{eqnarray}
\left.\left(u_x+\beta u\right)\right|_{x=0}=0,
\label{rbc}
\end{eqnarray}
for the classical NLS equation (\ref{nls}).
For $k=2$, we obtain the following boundary conditions
\begin{eqnarray}
\left.\left(u_x+\beta u\right)\right|_{x=0}=0,
~~\left.\left(u_{xxx}+\beta\left(u_{xx}-4|u|^2u\right)\right)\right|_{x=0}=0,
\label{rbcfnls}
\end{eqnarray}
for the fourth order NLS equation (\ref{fnls}).
We note that the boundary conditions (\ref{bceh}) for the $2k$th NLS equation ((\ref{nlsh}) with $n=2k$)) coincide with the ones found in \cite{WZ2022}.

{\bf Case II. New boundary conditions corresponding to $\varepsilon=-1$.} In this case, we may deduce from (\ref{abc}) that
\begin{eqnarray}
\begin{split}
\left.\tilde{c}_{2j-1}\right|_{x=0}=-\left.c_{2j-1}\right|_{x=0},
~~\left.\tilde{c}_{2j}\right|_{x=0}=\left.c_{2j}\right|_{x=0},
~~1\leq j\leq k,
\\
\left.\tilde{a}_{2j-1}\right|_{x=0}=-\left.a_{2j-1}\right|_{x=0},
~~\left.\tilde{a}_{2j}\right|_{x=0}=\left.a_{2j}\right|_{x=0},
~~1\leq j\leq k.
\end{split}
\label{acred2}
\end{eqnarray}
Using (\ref{acred2}), we find that, for $j$ being even numbers, equation (\ref{BTNLSa}) evaluated at $x=0$ becomes an identity when $\alpha=0$. For other equations in (\ref{BTNLSa}) and for equation (\ref{BTNLSb}) evaluated at $x=0$, we obtain the following new boundary conditions for the $2k$th NLS equation:
\begin{eqnarray}
\begin{split}
\left.\left(c_{2l+1}-2iua_{2l}-i\Omega_1 c_{2l}\right)\right|_{x=0}=0,~~1\leq l\leq k-1,
\\
\left.\left(u_t+2ua_{n}+\Omega_1 c_{n}\right)\right|_{x=0}=0,
\end{split}
\label{bceh2}
\end{eqnarray}
where $\Omega_1=\sqrt{\beta^2 - 4|u|^2}$.
The main difference with respect to the boundary conditions (\ref{bceh}) is that the boundary conditions (\ref{bceh2}) involve the time derivative of the potential.
In particular, for $k=1$ we obtain the following boundary condition
\begin{eqnarray}
\left.\left(u_t-2i|u|^2u+i\Omega_1u_x\right)\right|_{x=0}=0,
\label{tbcnls}
\end{eqnarray}
for the classical NLS equation (\ref{nls}).
For $k=2$, we obtain the following boundary conditions
\begin{eqnarray}
\begin{split}
\left.\left(u_{xx}+u_x\Omega_1\right)\right|_{x=0}=0,
\\
\left.\left(u_t+2iu\left(3|u|^4-|u_x|^2\right)-i\Omega_1\left(u_{xxx}+8|u|^2u_x+2u^2\bar{u}_x\right)\right)\right|_{x=0}=0,
\end{split}
\label{tbcfnls}
\end{eqnarray}
for the fourth order NLS equation (\ref{fnls}).

\begin{remark}
Despite that by using the bulk equation to eliminate $u_t$ the boundary condition (\ref{tbcnls}) coincides with the one appeared in \cite{Habibullin1991} (see the boundary condition (3.3) in \cite{Habibullin1991}), very little work was done to address the integrability and the solution methods to the boundary condition (\ref{tbcnls}), not to mention its higher order generalization (\ref{bceh2}). The study on these aspects will be done in this paper (see section 4.1 for the integrability and section 5.1 for a solution method).
\end{remark}

\begin{remark}
By dressing a Dirichlet boundary with an integrable defect, Zambon in \cite{Zambon2014} presented the following boundary condition
\begin{eqnarray}
\left.\left(iu_t-2u_{x}\sqrt{4b^2-|u|^2}-4(a^2+b^2)u+2 u|u|^2\right)\right|_{x=0}=0,
\label{tbc}
\end{eqnarray}
where $a$ and $b$ are two arbitrary real parameters, for the NLS equation. We note that the boundary condition (\ref{tbcnls}) is different from (\ref{tbc}), despite that they look similar in form. Indeed, the $K(\lambda)$ matrix characterizing the boundary (\ref{tbcnls}) is of order $1$ in $\lambda^{-1}$ (see section 4.1 in the following), while the $K(\lambda)$ matrix characterizing the boundary (\ref{tbc}) is of order $2$ in $\lambda^{-1}$ (see \cite{Zambon2014}).
\end{remark}

\subsection{Boundary conditions for the odd order NLS equations}

For $n=2k+1$ is odd, the corresponding $n$th NLS equation does not admit the reduction (\ref{red1}), but admits, instead, the following space-inverse and time-reverse reduction
\begin{eqnarray}
\tilde{u}(x,t)=\varepsilon u(-x,-t),~~\varepsilon=\pm 1.
\label{red2}
\end{eqnarray}
In this case, we may deduce that
\begin{eqnarray}
\begin{split}
\left.\tilde{c}_{2j+1}(x,t)\right|_{x=0}=\varepsilon\left.c_{2j+1}(x,-t)\right|_{x=0},
~~\left.\tilde{c}_{2j}(x,t)\right|_{x=0}=-\varepsilon\left.c_{2j}(x,-t)\right|_{x=0},
~~0\leq j\leq k,
\\
\left.\tilde{a}_{2j+1}(x,t)\right|_{x=0}=-\left.a_{2j+1}(x,-t)\right|_{x=0},
~~\left.\tilde{a}_{2j}(x,t)\right|_{x=0}=\left.a_{2j}(x,-t)\right|_{x=0},
~~0\leq j\leq k.
\end{split}
\label{acred3}
\end{eqnarray}
By using (\ref{acred3}), we obtain, from (\ref{BTNLS}) with the parameter $\alpha=0$, the following new boundary conditions at $x=0$,
\begin{eqnarray}
\begin{split}
(-1)^{j+1}\varepsilon\hat{c}_j-c_j-i\left(\varepsilon\hat{u}-u\right)
\left((-1)^{j-1}\hat{a}_{j-1}+a_{j-1}\right)
\\
+i\Omega_2\left((-1)^{j}\varepsilon\hat{c}_{j-1}+c_{j-1}\right)=0,
~~2\leq j\leq 2k+1,
\\
\varepsilon\hat{u}_t-u_t+\left(\varepsilon\hat{u}-u\right)\left(a_n-\hat{a}_n\right)
-\Omega_2\left(c_{n}+\varepsilon\hat{c}_{n}\right)=0,
\end{split}
\label{bconls}
\end{eqnarray}
where
\begin{eqnarray}
\hat{u}=u(x,-t),~~\hat{c}_j=c_j(x,-t),~~\hat{a}_j=a_j(x,-t),
~~
\Omega_2=\sqrt{\beta^2 - |u-\varepsilon\hat{u}|^2}.
\end{eqnarray}
The main novelty of the boundary conditions (\ref{bconls}) in comparison with known ones is that the boundaries (\ref{bconls}) involve the time reversal. In particular, for $k=1$, we obtain the following boundary conditions at $x=0$,
\begin{eqnarray}
\begin{split}
\varepsilon\hat{u}_x+u_x=&-\Omega_2\left(u+\varepsilon\hat{u}\right),
\\
u_{xx}-\varepsilon\hat{u}_{xx}=&\left(u+\varepsilon\hat{u}\right)\left(|\hat{u}|^2-|u|^2\right)
-\Omega_2\left(u_x-\varepsilon\hat{u}_x\right),
\\
u_t-\varepsilon\hat{u}_t=&\left(\varepsilon\hat{u}-u\right)
\left(u\bar{u}_x-u_x\bar{u}-\hat{u}\bar{\hat{u}}_x+\hat{u}_x\bar{\hat{u}}\right)
-\left(u_{xx}+2|u|^2u\right)\Omega_2
-\varepsilon\left(\hat{u}_{xx}+2|\hat{u}|^2\hat{u}\right)\Omega_2,
\end{split}
\label{bccmkdv}
\end{eqnarray}
for the complex mKdV equation (\ref{cmkdv}).

We note that we have set the parameter $\alpha$ appearing in (\ref{nlsdm}) to be zero for the boundary conditions discussed both in this subsection and in the above subsection. The reason for doing this will be explained in section 4.

%We will show the boundary conditions (\ref{bconls2}) are integrable and the associated boundary problems can be solved by using the DT method.

%{\bf Remark} We note that in \cite{} the authors also discussed the integrable BCs for the NLS hiera. However, in \cite{} only the higher order generalization of the Robin boundary condition , that is (), is obtained.
%The boundary condition () and () presented in this paper did not presented in \cite{}.
%No integrable BSc for the odd order NLS equations is obtained in the literature.

\section{Integrability of the boundary conditions}

In this section we develop the Sklyanin's approach to establish the integrability of the new boundary conditions (\ref{bceh2}) and (\ref{bconls}). The main difficulty in doing so is to find the $K(\lambda)$ matrices describing our new boundary conditions. Based on a connection between the time-part BT equations  and the equation characterising the boundary constraint in Sklyanin's approach, we give a method to construct the corresponding $K(\lambda)$ matrices directly from the BT matrices.

For fixing the idea, we focus on the problem on the positive half $x$-axis and we assume that the potentials and their derivatives satisfy Schwartz boundary conditions as $x$ going to infinity.
%we assume the fields and their derivatives decay to zero at infinity.

\subsection{Integrability of boundary conditions for even order NLS equations}

We consider the canonical Poisson brackets
\begin{eqnarray}
\left\{u(x,t),u(y,t)\right\}=\left\{\bar{u}(x,t),\bar{u}(y,t)\right\}=0,
~~
\left\{u(x,t),\bar{u}(y,t)\right\}=i\delta(x-y),
\label{cPB}
\end{eqnarray}
where $\delta(x-y)$ is the Dirac $\delta$-function.
Using the above Poisson brackets, one can deduce that the transition matrix
\begin{eqnarray}
T(x,y,t,\lambda)=\overset{\curvearrowleft}\exp \int_{y}^{x} U(\xi,t,\lambda)d\xi
\end{eqnarray}
satisfies the following well-known relation (see e.g. \cite{Faddeev2007})
\begin{eqnarray}
\left\{T_1(x,y,t,\lambda),T_2(x,y,t,\mu)\right\}=\left[r(\lambda-\mu),T_1(x,y,t,\lambda)T_2(x,y,t,\mu)\right],
\label{rmrelation}
\end{eqnarray}
where $T_1(x,y,t,\lambda)=T(x,y,t,\lambda)\otimes I$, $T_2(x,y,t,\mu)=I\otimes T(x,y,t,\mu)$, and the classical $r$-matrix $r$ is
\begin{eqnarray}
r(\lambda)=\frac{1}{\lambda}\left( \begin{array}{cccc}
1 & 0 & 0 & 0
\\
0 &  0 & 1 & 0
\\
0 & 1 & 0 & 0
\\
0 &  0 & 0 & 1
 \\ \end{array} \right).
 \label{rm}
\end{eqnarray}
For the $n$th NLS equation, the evolution of $T(x,y,t,\lambda)$ along $t$ is given by
\begin{eqnarray}
 \frac{\partial T(x,y,t,\lambda)}{\partial t}=V^{(n)}(x,t,\lambda)T(x,y,t,\lambda)
 -T(x,y,t,\lambda)V^{(n)}(y,t,\lambda),~~ n\geq 2.
 \label{Td}
\end{eqnarray}

For the $2k$th NLS equation, we introduce the following generalization of the monodromy matrix $T(\lambda)$ \cite{Sklyanin1987}
\begin{eqnarray}
\tau(\lambda)=T(t,\lambda)K(\lambda)T^{-1}(t,-\lambda),
\label{gm}
\end{eqnarray}
where $T(t,\lambda)=T(\infty,0,t,\lambda)$.
We assume that the $K(\lambda)$ matrix, in general, can depend on time.
Then one may deduce that (see e.g. \cite{Sklyanin1987,ACC2018}) integrable boundary conditions for the $2k$th NLS equation on the half-line are encoded into the $K(\lambda)$ matrices that satisfy both
\begin{eqnarray}
\left\{K_1(\lambda),K_2(\mu)\right\}=\left[r(\lambda-\mu),K_1(\lambda)K_2(\mu)\right]
+K_1(\lambda)r(\lambda+\mu)K_2(\mu)-K_2(\mu)r(\lambda+\mu)K_1(\lambda),
 \label{alga}
\end{eqnarray}
and
\begin{eqnarray}
 \frac{dK(\lambda)}{dt}=V^{(2k)}(0,t,\lambda)K(\lambda)-K(\lambda)V^{(2k)}(0,t,-\lambda).
 \label{algc}
\end{eqnarray}
Indeed, (\ref{algc}) implies that $\tr(\tau(\lambda))$ provides a generator for the conserved quantities,
while (\ref{alga}) ensures the Poisson commutativity of these conserved quantities: $\left\{\tr(\tau(\lambda)),\tr(\tau(\mu))\right\}=0$.

To establish the integrability of the $2k$th NLS equation in the presence of boundary conditions (\ref{bceh2}), we need to find the corresponding $K(\lambda)$ matrix. From (\ref{BT1b}) and (\ref{algc}), we obtain the following result which can be verified directly.
\begin{proposition}\label{proK}
Let the $2k$th NLS potentials $\tilde{u}$ and $u$ subject to a suitable reduction
(say the reduction (\ref{red1}) for example).
If, under this reduction, there exists a non-degenerate and time-independent matrix $\mathcal{C}(\lambda)$ such that
\begin{eqnarray}
\tilde{V}^{(2k)}(0,t,\lambda)=\mathcal{C}(\lambda)V^{(2k)}(0,t,-\lambda)\mathcal{C}^{-1}(\lambda),
\label{VP}
\end{eqnarray}
then
\begin{eqnarray}
K(\lambda)=B(0,t,\lambda)\mathcal{C}(\lambda),
\label{KB}
\end{eqnarray}
where $B(0,t,\lambda)$ stands for the BT matrix evaluated at $x=0$,
satisfies the boundary equation (\ref{algc}).
\end{proposition}

For the reduction (\ref{red1}) with $\varepsilon=1$, we find, by using (\ref{acred1}), that the $\mathcal{C}(\lambda)$ matrix corresponding to (\ref{VP}) is given by $\mathcal{C}=\diag(-1,1)$.
Recall that we have set $\alpha=0$. Then we recover the $K(\lambda)$ matrix (see \cite{Sklyanin1987}) for the higher order generalized Robin boundary conditions (\ref{bceh}),
\begin{eqnarray}
K(\lambda)=\left( \begin{array}{cc} -\lambda+i\beta & 0 \\
 0 &  \lambda+i\beta  \\ \end{array} \right).
 \label{K1}
\end{eqnarray}

For the reduction (\ref{red1}) with $\varepsilon=-1$, we find, by using (\ref{acred2}), that the $\mathcal{C}(\lambda)$ matrix corresponding to (\ref{VP}) is the identity matrix.
Then, by using (\ref{KB}) we obtain that the $K(\lambda)$ matrix for the boundary conditions (\ref{bceh2}) is given by
\begin{eqnarray}
K(\lambda)=\lambda\mathbf{I}+\left( \begin{array}{cc} -i\mathbf{\Omega}_1 & 2i\bar{\mathbf{u}}\\
 2i \mathbf{u} &  i\mathbf{\Omega}_1  \\ \end{array} \right),
 \label{K2}
\end{eqnarray}
where
\begin{eqnarray}
\mathbf{u}=\left.u(x,t)\right|_{x=0},
~~\mathbf{\Omega}_1=\sqrt{\beta^2 - 4|\mathbf{u}|^2}.
\label{mbuomg}
\end{eqnarray}

We note that equation (\ref{algc}) implies that $K(-\lambda,t)$ and $K^{-1}(\lambda,t)$ satisfy the same constraint equation.
This fact yields that the boundary $K(\lambda,t)$ matrix subjects to the following restriction
\begin{eqnarray}
K(\lambda,t)K(-\lambda,t)=f(\lambda)\mathbf{I},
\label{KC1}
\end{eqnarray}
where $f(\lambda)$ is some even function of $\lambda$. Our choice of the parameter $\alpha$ being zero in section 3  ensures that the resulting $K(\lambda,t)$ matrices obey (\ref{KC1}). Indeed, for both the $K(\lambda,t)$ matrices (\ref{K1}) and (\ref{K2}), we have
\begin{eqnarray}
K(\lambda,t)K(-\lambda,t)=-\left(\lambda^2+\beta^2\right)\mathbf{I}.
\label{KC12}
\end{eqnarray}
%For the Robin boundary condition, it has been shown that the property (\ref{KC12}) of the $K(\lambda)$ matrix (\ref{K1}) is important for solving the corresponding boundary problem (see e.g. \cite{Zhang2019}). For the new boundary condition (\ref{bceh2}), we will show in section 5 it is also the case.

Unlike (\ref{K1}), the $K(\lambda)$ matrix (\ref{K2}) is dynamical. Thus the corresponding Poisson bracket (\ref{alga}) is not automatically zero.
We need to check that the $K(\lambda,t)$ matrix (\ref{K2}) does match (\ref{alga}).
After straightforward calculation, we find that it is indeed the case and the Poisson algebra (\ref{alga}) is equivalent to the following Poisson bracket at boundary
\begin{eqnarray}
\left\{\mathbf{u},\bar{\mathbf{u}}\right\}=-i\mathbf{\Omega}_1.
\label{bpb1}
\end{eqnarray}
In fact, by using (\ref{K2}) and by extracting $(32)$-entry from (\ref{alga}), we obtain the boundary Poisson bracket (\ref{bpb1}). The Poisson brackets in other entries of (\ref{alga}) can be consistently deduced from (\ref{bpb1}). We note that the boundary Poisson bracket (\ref{bpb1}) is important for the integrability of the new boundary conditions (\ref{bceh2}), since it ensures the equality (\ref{alga}) and thus it guarantees the Poisson commutativity of the conserved quantities generated from the trace of the monodromy matrix (\ref{gm}).

Gathering the above arguments, we obtain the integrability of the $2k$th NLS equation in the presence of the boundary conditions (\ref{bceh2}).

\subsection{Integrability of the boundary conditions for odd order NLS equations}

To establish the integrability of the $(2k+1)$th NLS equation with the boundary conditions (\ref{bconls}), we need to further develop the Sklyanin's method. In this case, we introduce, instead of (\ref{gm}), the following generalization of the monodromy matrix,
\begin{eqnarray}
\tau(\lambda)=T(t,\lambda)K(\lambda)T^{-1}(-t,-\lambda),
\label{gm2}
\end{eqnarray}
where, as before, $T(t,\lambda)=T(\infty,0,t,\lambda)$.

\begin{proposition}
Let $K(\lambda)$ satisfy the relation
\begin{eqnarray}
\left\{K_1(\lambda),K_2(\mu)\right\}=\left[r(\lambda-\mu),K_1(\lambda)K_2(\mu)\right],
 \label{alga2}
\end{eqnarray}
then the quantities $\tr(\tau(\lambda))$ are in involution: $\left\{\tr(\tau(\lambda)),\tr(\tau(\mu))\right\}=0$.
\end{proposition}
{\bf Proof}
Using Poisson bracket (\ref{rmrelation}), we find
\begin{eqnarray}
\left\{T_1(-t,-\lambda),T_2(-t,-\mu)\right\}=-\left[r(\lambda-\mu),T_1(-t,-\lambda)T_2(-t,-\mu)\right],
\end{eqnarray}
and thus
\begin{eqnarray}
\begin{split}
&\left\{T^{-1}_1(-t,-\lambda),T^{-1}_2(-t,-\mu)\right\}
\\
=&T^{-1}_2(-t,-\mu)T^{-1}_1(-t,-\lambda)\left\{T_1(-t,-\lambda),T_2(-t,-\mu)\right\}T^{-1}_1(-t,-\lambda)T^{-1}_2(-t,-\mu)
\\
=&\left[r(\lambda-\mu),T^{-1}_1(-t,-\lambda)T^{-1}_2(-t,-\mu)\right].
\end{split}
\label{rmrelation1}
\end{eqnarray}
Using $\left\{T_1(t,\lambda),T_2(-t,-\mu)\right\}=\left\{T_1(-t,-\lambda),T_2(t,\mu)\right\}=0$,
we have
\begin{eqnarray}
\begin{split}
\left\{T_1(t,\lambda),T^{-1}_2(-t,-\mu)\right\}
=-T^{-1}_2(-t,-\mu)\left\{T_1(t,\lambda),T_2(-t,-\mu)\right\}T^{-1}_2(-t,-\mu)=0,
\\
\left\{T^{-1}_1(-t,-\lambda),T_2(t,\mu)\right\}
=-T^{-1}_1(-t,-\lambda)\left\{T_1(-t,-\lambda),T_2(t,\mu)\right\}T^{-1}_1(-t,-\lambda)
=0.
\end{split}
\label{rmrelation2}
\end{eqnarray}
After straightforward calculations using (\ref{rmrelation}), (\ref{rmrelation1}) and (\ref{rmrelation2}),
we obtain
\begin{eqnarray}
\begin{split}
&\left\{\tau_1(\lambda),\tau_2(\mu)\right\}
\\
=&\left[r(\lambda-\mu),\tau_1(\lambda)\tau_2(\mu)\right]
\\
&+T_1(t,\lambda)T_2(t,\mu)
\left(\left\{K_1(\lambda),K_2(\mu)\right\}-\left[r(\lambda-\mu),K_1(\lambda)K_2(\mu)\right]\right)
T^{-1}_1(-t,-\lambda)T^{-1}_2(-t,-\mu).
\end{split}
 \label{taupb}
\end{eqnarray}
Using the above relation, we complete the proof. \QEDB

\begin{proposition}
Consider the $(2k+1)$th NLS equation (namely, equation (\ref{nlsh}) with $n=2k+1$) on the half-line.
Let the boundary conditions at $x=0$ be determined by
\begin{eqnarray}
 \frac{dK(\lambda)}{dt}=V^{(2k+1)}(0,t,\lambda)K(\lambda)+K(\lambda)V^{(2k+1)}(0,-t,-\lambda).
 \label{algc2}
\end{eqnarray}
Then $\frac{d\tr(\tau(\lambda))}{dt}=0$.
\end{proposition}
{\bf Proof}
By using the formula (\ref{Td}), we obtain
\begin{eqnarray}
 \frac{d T(t,\lambda)}{dt}=&V^{(2k+1)}(\infty,t,\lambda)T(t,\lambda)
 -T(t,\lambda)V^{(2k+1)}(0,t,\lambda),
 \label{Td2}
\end{eqnarray}
 and
\begin{eqnarray}
 \begin{split}
 \frac{d T^{-1}(-t,-\lambda)}{dt}
 =&-T^{-1}(-t,-\lambda)\frac{d T(-t,-\lambda)}{dt}T^{-1}(-t,-\lambda)
 \\
 =&T^{-1}(-t,-\lambda)V^{(2k+1)}(\infty,-t,-\lambda)-V^{(2k+1)}(0,-t,-\lambda)T^{-1}(-t,-\lambda).
 \end{split}
\label{Td3}
\end{eqnarray}
Using (\ref{Td2}) and (\ref{Td3}), we find
\begin{eqnarray}
 \begin{split}
 \frac{d\tau(\lambda)}{dt}
 =&V^{(2k+1)}(\infty,t,\lambda)T(t,\lambda)K(\lambda)T^{-1}(-t,-\lambda)
 +T(t,\lambda)K(\lambda)T^{-1}(-t,-\lambda)V^{(2k+1)}(\infty,-t,-\lambda)
 \\
 &+T(t,\lambda)
 \left(\frac{dK(\lambda)}{dt}-V^{(2k+1)}(0,t,\lambda)K(\lambda)-K(\lambda)V^{(2k+1)}(0,-t,-\lambda)\right)
 T^{-1}(-t,-\lambda).
 \end{split}
\label{taud}
\end{eqnarray}
For the problem with vanishing boundary conditions at infinity, we have
\begin{eqnarray}
 V^{(2k+1)}(\infty,t,\lambda)=-V^{(2k+1)}(\infty,-t,-\lambda)
 =\diag\left(\frac{i}{2}\lambda^{(2k+1)},-\frac{i}{2}\lambda^{(2k+1)}\right).
\label{vinf}
\end{eqnarray}
Inserting (\ref{vinf}) into (\ref{taud}), we complete the proof. \QEDB

To prove the integrability of the $(2k+1)$th NLS equation in the presence of boundary conditions (\ref{bconls}), we need to find the corresponding $K(\lambda)$ matrix that matches both equations (\ref{alga2}) and (\ref{algc2}).
Observing (\ref{BT1b}) and (\ref{algc2}), we obtain the following conclusion.
\begin{proposition}\label{proK2}
If, under a suitable reduction (say the reduction (\ref{red2}) for example), there exists a non-degenerate and time-independent matrix $\mathcal{C}(\lambda)$ such that
\begin{eqnarray}
\tilde{V}^{(2k+1)}(0,t,\lambda)=-\mathcal{C}(\lambda)V^{(2k+1)}(0,-t,-\lambda)\mathcal{C}^{-1}(\lambda),
\label{VP2}
\end{eqnarray}
then
$K(\lambda)=B(0,t,\lambda)\mathcal{C}(\lambda)$
satisfies the boundary equation (\ref{algc2}).
\end{proposition}

For the reduction (\ref{red2}), we find, by using (\ref{acred3}), that the $\mathcal{C}(\lambda)$ matrix corresponding to (\ref{VP2}) can be taken as $\mathcal{C}=\diag(-\varepsilon,1)$.
Recall that we have set the parameter $\alpha$ appearing in (\ref{nlsdm}) to be zero for the boundary conditions (\ref{bconls}). Then by applying proposition \ref{proK2} we obtain
\begin{eqnarray}
K(\lambda)=\lambda\left( \begin{array}{cc} -\varepsilon & 0 \\ 0 & 1\\ \end{array} \right)
+\left( \begin{array}{cc} i\varepsilon\mathbf{\Omega}_2 & i\left(\bar{\mathbf{u}}-\varepsilon\bar{\hat{\mathbf{u}}}\right)\\
 i\left( \hat{\mathbf{u}}-\varepsilon\mathbf{u} \right)&  i\mathbf{\Omega}_2 \\
\end{array} \right),
 \label{K3}
\end{eqnarray}
where $\mathbf{u}$ is defined by (\ref{mbuomg}), and
\begin{eqnarray}
\hat{\mathbf{u}}=\left.u(x,-t)\right|_{x=0},
~~\mathbf{\Omega}_2=\sqrt{\beta^2 - |\mathbf{u}-\varepsilon\hat{\mathbf{u}}|^2}.
\label{mbuomgh}
\end{eqnarray}
We can verify directly that the $K(\lambda)$ matrix (\ref{K3}) does describe the boundary conditions (\ref{bconls}), i.e. equation (\ref{algc2}) with $K(\lambda)$ matrix given by (\ref{K3}) does produce the boundary conditions (\ref{bconls}).
We note that the boundary matrix (\ref{K3}) satisfies
\begin{eqnarray}
K(\lambda,-t)K(-\lambda,t)=-(\lambda^2+\beta^2)\mathbf{I}.
\label{KC21}
\end{eqnarray}
This property is important to solve the associated boundary problem via boundary dressing method (see relation (\ref{K0C}) employed in the proof of the proposition \ref{solution} in section 5).

Using (\ref{K3}), we find that the Poisson algebra (\ref{alga2}) is equivalent to the following boundary Poisson bracket
\begin{eqnarray}
\left\{\mathfrak{u},\bar{\mathfrak{u}}\right\}=-2i\mathbf{\Omega}_2,
\label{bpb2}
\end{eqnarray}
where $\mathfrak{u}= \varepsilon\hat{\mathbf{u}}-\mathbf{u}$.
In fact, extracting $(32)$-entry from (\ref{alga2}) yields the boundary Poisson bracket (\ref{bpb2}). The Poisson brackets appearing in other entries of (\ref{alga2}) can be consistently deduced from (\ref{bpb2}). To sum up, we obtain that the Poisson algebra (\ref{alga2}) becomes an identity provided that the boundary Poisson bracket is defined by (\ref{bpb2}). Thus, integrability of the $(2k+1)$th NLS equation with the boundary conditions (\ref{bconls}) is obtained.

\subsection{Conserved quantities}

We now derive explicit forms of the Poisson commuting conserved quantities for the $2k$th NLS equation in the presence of the new boundary conditions (\ref{bceh2}) and for the $(2k+1)$th NLS equation in the presence of the new boundary conditions (\ref{bconls}) by investigating the large $\lambda$ expansions of the trace of the corresponding monodromy matrices.

Let us first focus on the boundary conditions (\ref{bconls}).
For $|\lambda|\rightarrow \infty$, the transition matrix $T$ can be expanded as \cite{Faddeev2007}
\begin{eqnarray}
T(x,y,t,\lambda)=\left(\mathbf{I}+W(x,t,\lambda)\right)\exp\left[Z(x,y,t,\lambda)\right]
\left(\mathbf{I}+W(y,t,\lambda)\right)^{-1},
\label{texp}
\end{eqnarray}
where $W$ is an off-diagonal matrix and $Z$ is a diagonal matrix.
Inserting (\ref{texp}) into (\ref{lpx}), one may obtain that the elements of $W$ and $Z$ have the following asymptotic representations
\begin{eqnarray}
\begin{split}
W_{21}(x,t,\lambda)=\sum_{n=1}^{\infty}\frac{\omega_n(x,t)}{(i\lambda)^n}, ~~W_{12}(x,t,\lambda)=-\overline{W_{21}(x,t,\bar{\lambda})},
\\
Z_{22}(x,y,t,\lambda)=\frac{i}{2}\lambda(x-y)+\int_{y}^{x}u(z,t)W_{12}(z,t,\lambda)dz,
~~Z_{11}(x,y,t,\lambda)=\overline{Z_{22}(x,y,t,\bar{\lambda})},
\end{split}
\label{WZexp}
\end{eqnarray}
where the series coefficients $\omega_n$ are given recursively by
\begin{eqnarray}
\begin{split}
\omega_1=-u, ~~\omega_2=-u_x,
~~
\omega_{n+1}=\left(\omega_n\right)_x-\bar{u}\sum_{k=1}^{n-1}\omega_k\omega_{n-k}.
\end{split}
\label{omgn}
\end{eqnarray}

Using (\ref{texp}), the trace of the monodromy matrix (\ref{gm2}) can be expressed as
\begin{eqnarray}
\tr\tau(\lambda)=\tr\left(\exp\left(Z-\hat{Z}\right)\left(\mathbf{I}+W(0)\right)^{-1}K(\lambda)
\left(\mathbf{I}+\hat{W}(0)\right)\right),
\end{eqnarray}
where $Z\equiv Z(\infty,0,t,\lambda)$, $\hat{Z}\equiv Z(\infty,0,-t,-\lambda)$, $W\equiv W(0,t,\lambda)$ and $\hat{W}\equiv W(0,-t,-\lambda)$.
The non-trivial part in the expansion of $\tr\tau(\lambda)$ for $i\lambda\rightarrow\infty$ comes from the $Z_{22}$ terms (the leading order in $Z_{11}(x,y,t,\lambda)$ is $-\frac{i}{2}\lambda(x-y)$, thus the contribution from $\exp \left[Z_{11}(\infty,0,t,\lambda)\right]$ vanishes as $i\lambda\rightarrow\infty$). Therefore, we only need to derive the asymptotic expansion of the $(22)$-element of $\left(\mathbf{I}+W(0)\right)^{-1}K(\lambda)\left(\mathbf{I}+\hat{W}(0)\right)$ as $i\lambda\rightarrow\infty$. Let us denote the $(22)$-element of $\left(\mathbf{I}+W(0)\right)^{-1}K(\lambda)\left(\mathbf{I}+\hat{W}(0)\right)$ by $G\left(\lambda\right)$ for short.
Recall that the $K(\lambda)$ matrix for the boundary conditions (\ref{bconls}) is given by (\ref{K3}).
To compute the expansion of $G\left(\lambda\right)$, it is convenient for us to multiply the $K(\lambda)$ matrix (\ref{K3}) by the factor $\lambda^{-1}$ (doing so has no effect on our boundary conditions, since the $K$ matrices are allowed up to a constant factor).
Then, after straightforward computation we obtain the expansion
\begin{eqnarray}
G\left(\lambda\right)=1+\sum_{n=1}^{\infty}\frac{G_n}{(i\lambda)^n},
\label{Gexp}
\end{eqnarray}
where the series coefficients $G_n$ are given recursively by
\begin{eqnarray}
\begin{split}
G_1=&-\Omega_2, ~~G_2=\varepsilon u \bar{\hat{u}}-|\hat{u}|^2,
\\
G_{n+1}=&\sum_{m=1}^{n-1}G_m\left(\sum_{l=1}^{n-m}(-1)^{n-m-l}\omega_l\bar{\omega}_{n+1-m-l}\right)
+\sum_{l=1}^{n}(-1)^{n-l}\omega_l\bar{\omega}_{n+1-l}
\\&
-(\varepsilon\bar{\hat{u}}-\bar{u})\omega_n
+(\hat{u}-\varepsilon u)\bar{\hat{\omega}}_n
-\varepsilon\sum_{l=1}^{n}\omega_l\bar{\hat{\omega}}_{n+1-l}
-\varepsilon\Omega_2\sum_{l=1}^{n-1}\omega_l\bar{\hat{\omega}}_{n-l},
\end{split}
\label{Gn}
\end{eqnarray}
evaluated at $x=0$. In (\ref{Gn}), we have used $\hat{\omega}_n$ to denote $\omega_n(x,-t)$ (recall that $\omega_n\equiv\omega_n(x,t)$ is defined by (\ref{omgn})).

Gathering all the above information, we finally obtain from $\ln\left(\tr\tau(\lambda)\right)$ the following generating function of the conserved quantities for the $(2k+1)$th NLS equation in the presence of the boundary conditions (\ref{bconls}):
\begin{eqnarray}
I(\lambda)=\int_{0}^{\infty}\left[u(x,t)W_{12}(x,t,\lambda)-u(x,-t)W_{12}(x,-t,-\lambda)\right]dx
+\ln\left(G(\lambda)\right),
\label{GCQ1}
\end{eqnarray}
where $W_{12}(\lambda)$ and $G(\lambda)$ are defined by expansions (\ref{WZexp}) and (\ref{Gexp}), respectively, with the expansion coefficients $\omega_n$ and $G_n$ given recursively by (\ref{omgn}) and (\ref{Gn}).
By expanding the logarithmic function $\ln\left(G(\lambda)\right)$ in $\lambda^{-1}$, we can extract explicit forms of the conserved quantities order by order. For example, the first three members are given by
\begin{eqnarray}
\begin{split}
I_1=\int_{0}^{\infty}\left(|u|^2+|\hat{u}|^2\right)dx+\left.\Omega_2\right|_{x=0},
\\
I_2=i\int_{0}^{\infty}\left(u\bar{u}_x-\bar{u}u_x+\bar{\hat{u}}\hat{u}_x-\hat{u}\bar{\hat{u}}_x\right)dx
+i\varepsilon\left.\left(u\bar{\hat{u}}-\bar{u}\hat{u}\right)\right|_{x=0},
\\
I_3=\int_{0}^{\infty}\left(|u|^4+|\hat{u}|^4-|u_x|^2-|\hat{u}_x|^2\right)dx
+\left.\left(\frac{(\Omega_2)^3}{3}+\Omega_2\left(|u|^2+|\hat{u}|^2\right)\right)\right|_{x=0}.
\end{split}
\end{eqnarray}

The conserved quantities for the $2k$th NLS equation with the boundary conditions (\ref{bceh2}) can be derived via a similar manner.
In this case, the trace of the corresponding monodromy matrix (\ref{gm}) can be expressed as
\begin{eqnarray}
\tr\tau(\lambda)=\tr\left(\exp\left(Z(\infty,0,t,\lambda)-Z(\infty,0,t,-\lambda)\right)\left(\mathbf{I}+W(0,t,\lambda)\right)^{-1}K(\lambda)
\left(\mathbf{I}+W(0,t,-\lambda)\right)\right).
\label{gmexp2}
\end{eqnarray}
Let us denote the $(22)$-element of $\left(\mathbf{I}+W(0,t,\lambda)\right)^{-1}K(\lambda)\left(\mathbf{I}+W(0,t,-\lambda)\right)$ by $F\left(\lambda\right)$ for short.
Recall that the $K(\lambda)$ matrix for the boundary conditions (\ref{bceh2}) is given by (\ref{K2}).
As before, we multiply the $K(\lambda)$ matrix (\ref{K2}) by the factor $\lambda^{-1}$ for convenience sake.
Then, we obtain the following expansion
\begin{eqnarray}
F\left(\lambda\right)=1+\sum_{n=1}^{\infty}\frac{F_n}{(i\lambda)^n},
\label{Fexp}
\end{eqnarray}
where the series coefficients $F_n$ are given recursively by
\begin{eqnarray}
\begin{split}
F_1=&-\Omega_1, ~~F_2=-|u|^2,~~F_3=-2u\bar{u}_x,
\\
F_{n+1}=&\sum_{m=1}^{n-1}F_m\left(\sum_{l=1}^{n-m}(-1)^{n-m-l}\omega_l\bar{\omega}_{n+1-m-l}\right)
+\sum_{l=1}^{n}(-1)^{n-l}\omega_l\bar{\omega}_{n+1-l}
\\
&+2\bar{u}\omega_n
+2 u\bar{\omega}_n
+\sum_{l=1}^{n}\omega_l\bar{\omega}_{n+1-l}
+\Omega_1\sum_{l=1}^{n-1}\omega_l\bar{\omega}_{n-l},
\end{split}
\label{Fn}
\end{eqnarray}
evaluated at $x=0$.
From $\ln\left(\tr\tau(\lambda)\right)$, we obtain the following generating function of the conserved quantities for the $2k$th NLS equation with the boundary conditions (\ref{bceh2}):
\begin{eqnarray}
I(\lambda)=\int_{0}^{\infty}u(x,t)\left[W_{12}(x,t,\lambda)-W_{12}(x,t,-\lambda)\right]dx
+\ln\left(F(\lambda)\right),
\label{GCQ2}
\end{eqnarray}
where $W_{12}(\lambda)$ and $F(\lambda)$ are defined by expansions (\ref{WZexp}) and (\ref{Fexp}), respectively, with the expansion coefficients $\omega_n$ and $F_n$ given recursively by (\ref{omgn}) and (\ref{Fn}).
By expanding the logarithmic function $\ln\left(F(\lambda)\right)$ in $\lambda^{-1}$, we can extract explicit forms of the conserved quantities order by order. For example, the first two nontrivial conserved quantities are given by
\begin{eqnarray}
\begin{split}
I_1=2\int_{0}^{\infty}|u|^2dx+\left.\Omega_1\right|_{x=0},
\\
I_3=\int_{0}^{\infty}\left(|u|^4-|u_x|^2\right)dx
+\left.\left(\frac{(\Omega_1)^3}{6}+\Omega_1|u|^2\right)\right|_{x=0}.
\end{split}
\end{eqnarray}

We note that the main difference between the conserved quantities for the boundary conditions (\ref{bceh2}) and the ones for (\ref{bconls}) is that the even conserved quantities become trivial for the former case while they survive for the latter case.

\section{Dressing the boundaries}

In this section, we develop further the boundary dressing method introduced in \cite{Zhang2019,ZZ2021,WZ2022} and developed in \cite{Gruner2020} to construct soliton solutions for the $2k$th NLS equation in the presence of the boundary conditions (\ref{bceh2}) and for the $(2k+1)$th NLS equation in the presence of the boundary conditions (\ref{bconls}). The key idea in this method is to choose suitable DT that preserves the boundary constraint of the system at each step of the dressing. For convenience sake, we choose $u[0]$ to be zero seed solution. We obtain the following results.

\begin{proposition}\label{solution2}
Consider the $2k$th NLS equation (namely, (\ref{nlsh}) with $n=2k$) on the half-line subject to the boundary conditions (\ref{bceh2}). Let $u[0]$ be zero seed solution and $\left\{\psi_j(\lambda_j),\check{\psi}_j(\check{\lambda}_j)\right\}$, $j=1,\cdots,N$, be $N$ paired special solutions of the undressed Lax system such that
\begin{eqnarray}
\left.\check{\psi}_j(x,t,\check{\lambda}_j)\right|_{x=0}=\left.K_0(\check{\lambda}_j)\psi_j(x,t,\lambda_j)\right|_{x=0},
~~\check{\lambda}_j=-\lambda_j, ~~\check{\lambda}_j\neq\lambda_k,
\label{psirel2}
\end{eqnarray}
where $\lambda_j\in \mathbb{C}\setminus\left(\mathbb{R}\cup i\mathbb{R}\right)$, $j=1,\cdots,N$, are $N$ distinct spectral parameters, and
\begin{eqnarray}
K_0(\lambda)=\diag\left(\lambda-i|\beta|,\lambda+i|\beta|\right).
\label{K02}
\end{eqnarray}
Let $D[2N]$ be the $2N$-fold DT constructed by using $\left\{\psi_1,\check{\psi}_1,\cdots,\psi_N,\check{\psi}_N\right\}$ and the corresponding spectral parameters $\left\{\lambda_1,\check{\lambda}_1,\cdots,\lambda_N,\check{\lambda}_N\right\}$.
Then so-constructed DT yields the solution $u[2N]$ to the $2k$th NLS equation on the half-line that satisfies the boundary conditions (\ref{bceh2}). We denote by $\check{u}[N]$ such a solution $u[2N]$.
\end{proposition}

\begin{proposition}\label{solution}
Consider the $(2k+1)$th NLS equation (namely, (\ref{nlsh}) with $n=2k+1$) on the half-line subject to the boundary conditions (\ref{bconls}). Let $u[0]$ be zero seed solution and $\left\{\psi_j(\lambda_j),\check{\psi}_j(\check{\lambda}_j)\right\}$, $j=1,\cdots,N$, be $N$ paired special solutions of the undressed Lax system such that
\begin{eqnarray}
\left.\check{\psi}_j(x,t,\check{\lambda}_j)\right|_{x=0}=\left.K_0(\check{\lambda}_j)\psi_j(x,-t,\lambda_j)\right|_{x=0},
~~\check{\lambda}_j=-\lambda_j, ~~\check{\lambda}_j\neq\lambda_k,
\label{psirel}
\end{eqnarray}
where $\lambda_j\in \mathbb{C}\setminus\left(\mathbb{R}\cup i\mathbb{R}\right)$, $j=1,\cdots,N$, are $N$ distinct spectral parameters, and
\begin{eqnarray}
K_0(\lambda)=\diag\left(\varepsilon\left(-\lambda+i|\beta|\right),\lambda+i|\beta|\right).
\label{K0}
\end{eqnarray}
Let $D[2N]$ be the $2N$-fold DT constructed by using $\left\{\psi_1,\check{\psi}_1,\cdots,\psi_N,\check{\psi}_N\right\}$ and the corresponding spectral parameters $\left\{\lambda_1,\check{\lambda}_1,\cdots,\lambda_N,\check{\lambda}_N\right\}$.
Then so-constructed DT yields the solution $u[2N]$ to the $(2k+1)$th NLS equation on the half-line that satisfies the boundary conditions (\ref{bconls}).
\end{proposition}

For economy of presentation, here we only give the proof for proposition \ref{solution}, the proof for proposition \ref{solution2} can be performed analogously.

{\bf Proof of proposition \ref{solution}}. To show that $u[2N]$ satisfies the boundary conditions (\ref{bconls}), we need to prove the relation
\begin{eqnarray}
 \frac{dK_{2N}(\lambda)}{dt}=V^{(2k+1)}_{[2N]}(0,t,\lambda)K_{2N}(\lambda)+K_{2N}(\lambda)V^{(2k+1)}_{[2N]}(0,-t,-\lambda),
 \label{algcdress}
\end{eqnarray}
where $K_{2N}(\lambda)$ is defined by (\ref{K3}) with $\mathbf{u}$ replaced by $\mathbf{u}[2N]\equiv\left.u[2N](x,t)\right|_{x=0}$, that is
\begin{eqnarray}
K_{2N}(\lambda)=\lambda\left( \begin{array}{cc} -\varepsilon & 0 \\ 0 & 1\\ \end{array} \right)
+\left( \begin{array}{cc} i\varepsilon \mathbf{\Omega}[2N] & i\left(\bar{\mathbf{u}}[2N]-\varepsilon\bar{\hat{\mathbf{u}}}[2N]\right)\\
 i\left( \hat{\mathbf{u}}[2N]-\varepsilon\mathbf{u}[2N] \right) &  i\mathbf{\Omega}[2N] \\
\end{array} \right),
 \label{KN}
\end{eqnarray}
with $\mathbf{u}[2N]=\left.u[2N](x,t)\right|_{x=0}$, $\hat{\mathbf{u}}[2N]=\left.u[2N](x,-t)\right|_{x=0}$ and $\mathbf{\Omega}[2N]=\sqrt{\beta^2 - |\mathbf{u}[2N]-\varepsilon\hat{\mathbf{u}}[2N]|^2}$.
By using
\begin{eqnarray}
\left(D[2N](x,t,\lambda)\right)_t=V^{(2k+1)}_{[2N]}(x,t,\lambda)D[2N](x,t,\lambda)
-D[2N](x,t,\lambda)V^{(2k+1)}_{[0]}(x,t,\lambda),
\end{eqnarray}
we can verify directly that if the following equality holds
\begin{eqnarray}
\left.D[2N](x,t,\lambda)K_0(\lambda)\right|_{x=0}=\left.K_{2N}(\lambda)D[2N](x,-t,-\lambda)\right|_{x=0},
\label{Drel}
\end{eqnarray}
then (\ref{algcdress}) holds.
Our strategy for proving (\ref{Drel}) is constituted of the following two steps: first show that there does exist a matrix $K_{2N}$ satisfying (\ref{Drel}), then show that such a matrix is exactly (\ref{KN}).

We now perform the first step. It is easy to see the unit vectors $\psi_0=(1,0)^T$ and $\varphi_0=(0,1)^T$ are kernel vectors of $K_0(\lambda)$ at $\lambda=\lambda_0\equiv i|\beta|$ and at $\lambda=\bar{\lambda}_0\equiv -i|\beta|$ respectively, i.e. $K_0(\lambda_0)\psi_0=0$, $K_0(\bar{\lambda}_0)\varphi_0=0$.
We construct the matrix $K_{2N}(\lambda)$ as $K_{2N}(\lambda)=\lambda\diag(-\varepsilon,1)+K^{(0)}_{2N}$, where $K^{(0)}_{2N}$ is defined by the solution of the following system of linear algebraic equations
\begin{eqnarray}
K^{(0)}_{2N}\left(\tilde{\psi}_0,\tilde{\varphi}_0\right)
=-\left(\lambda_0\mathcal{C}\tilde{\psi}_0,\bar{\lambda}_0\mathcal{C}\tilde{\varphi}_0\right),
\label{kn0}
\end{eqnarray}
where $\tilde{\psi}_0=\left.D[2N](x,-t,-\lambda_0)\right|_{x=0}\psi_0$, $\tilde{\varphi}_0=\left.D[2N](x,-t,-\bar{\lambda}_0)\right|_{x=0}\varphi_0$ and $\mathcal{C}=\diag\left(-\varepsilon,1\right)$.
We note that the symmetry relations (see (\ref{symdt})) of the DT matrix $D[2N]$ implies that $\det\left(\tilde{\psi}_0,\tilde{\varphi}_0\right)\neq 0$. Thus system (\ref{kn0}) determines $K^{(0)}_{2N}$ uniquely.
We need to show that the so-constructed $K_{2N}(\lambda)$ matrix satisfies the equality (\ref{Drel}). The left hand side and the right hand side of (\ref{Drel}) are matrix polynomials of degree $2N+1$ in $\lambda$. We write them respectively as
\begin{subequations}
\begin{eqnarray}
L(\lambda)=\lambda^{2N+1}L_{2N+1}+\lambda^{2N}L_{2N}+\cdots+\lambda L_1+L_0,
\label{l}
\\
R(\lambda)=\lambda^{2N+1}R_{2N+1}+\lambda^{2N}R_{2N}+\cdots+\lambda R_1+R_0.
\label{r}
\end{eqnarray}
\label{lr}
\end{subequations}
It is easy to see $L_{2N+1}=R_{2N+1}=\diag(-\varepsilon,1)$. To prove that the remaining $2N+1$ matrix coefficients of (\ref{l}) equal to those of (\ref{r}), we show that $L(\lambda)$ and $R(\lambda)$ share the same $4N+2$ zeros and the associated kernel vectors. First, the construction of $K_{2N}(\lambda)$ implies the following relations
\begin{eqnarray}
\begin{split}
\left.D[2N](0,t,\lambda)K_0(\lambda)\right|_{\lambda=\lambda_0}\psi_0
=\left.K_{2N}(\lambda)D[2N](0,-t,-\lambda)\right|_{\lambda=\lambda_0}\psi_0=0,
\\
\left.D[2N](0,t,\lambda)K_0(\lambda)\right|_{\lambda=\bar{\lambda}_0}\varphi_0
=\left.K_{2N}(\lambda)D[2N](0,-t,-\lambda)\right|_{\lambda=\bar{\lambda}_0}\varphi_0=0.
\end{split}
\label{Drel0}
\end{eqnarray}
Moreover, we find the following equalities at $x=0$,
\begin{eqnarray}
\begin{split}
\left.R(\lambda)\right|_{\lambda=\lambda_j}\check{\psi}_j(-t)=0,
~~\left.R(\lambda)\right|_{\lambda=\check{\lambda}_j}\psi_j(-t)=0, ~~1\leq j\leq N,
\\
\left.R(\lambda)\right|_{\lambda=\bar{\lambda}_j}\check{\varphi}_j(-t)=0,
~~\left.R(\lambda)\right|_{\lambda=\bar{\check{\lambda}}_j}\varphi_j(-t)=0, ~~1\leq j\leq N,
\end{split}
\label{rkernel}
\end{eqnarray}
and
\begin{eqnarray}
\begin{split}
\left.L(\lambda)\right|_{\lambda=\lambda_j}\check{\psi}_j(-t)=0,
~~\left.L(\lambda)\right|_{\lambda=\check{\lambda}_j}\psi_j(-t)=0, ~~1\leq j\leq N,
\\
\left.L(\lambda)\right|_{\lambda=\bar{\lambda}_j}\check{\varphi}_j(-t)=0,
~~\left.L(\lambda)\right|_{\lambda=\bar{\check{\lambda}}_j}\varphi_j(-t)=0, ~~1\leq j\leq N,
\end{split}
\label{lkernel}
\end{eqnarray}
where $\varphi_j=\left(-\bar{\nu}_j,\bar{\mu}_j\right)^T$ is the orthogonal vector of $\psi_j=\left(\mu_j,\nu_j\right)^T$, and similarly for the notation $\check{\varphi}_j$.
Indeed, the equalities (\ref{rkernel}) follow from the definition of the DT matrix $D[2N]$.
The equalities (\ref{lkernel}) result from the assumption (\ref{psirel}) together with the following equality
\begin{eqnarray}
K_0(\lambda)K_0(-\lambda)=-(\lambda^2+\beta^2)\mathbf{I}.
\label{K0C}
\end{eqnarray}
Note that the $4N+2$ zeros $\left\{\lambda_0,\bar{\lambda}_0,\lambda_j,\bar{\lambda}_j,\check{\lambda}_j,\bar{\check{\lambda}}_j\right\}$, $1\leq j\leq N$, are distinct and the associated kernel vectors are linearly independent. Thus the $K_{2N}(\lambda)$ matrix constructed in this step does satisfy (\ref{Drel}).

Next we perform the second step. More precisely, we will show that the above constructed $K_{2N}(\lambda)$ is exactly (\ref{KN}). The equality $L_{2N}=R_{2N}$ yields
\begin{eqnarray}
K^{(0)}_{2N}=K^{(0)}_0+\left.\Sigma_1(x,t)\right|_{x=0}\mathcal{C}
+\mathcal{C}\left.\Sigma_1(x,-t)\right|_{x=0},
\label{KN0M}
\end{eqnarray}
where $K^{(0)}_0=\diag\left(i\varepsilon |\beta|,i|\beta|\right)$, $\mathcal{C}=\diag\left(-\varepsilon,1\right)$, and $\Sigma_1(x,t)$ is the matrix coefficient of $\lambda^{2N-1}$ of $D[2N](x,t,\lambda)$. Notice that
\begin{eqnarray}
\left(\Sigma_1(x,t)\right)_{21}=-i\left(u[0]-u[2N]\right),
~~\left(\Sigma_1(x,t)\right)_{12}=-i\left(\bar{u}[0]-\bar{u}[2N]\right),
\label{sigma1}
\end{eqnarray}
where $\left(\Sigma_1(x,t)\right)_{jk}$, $j,k=1,2$, stand for the $jk$-elements of the matrix $\Sigma_1(x,t)$. We obtain from (\ref{KN0M}) that the off-diagonal elements of $K^{(0)}_{2N}$ are given by
\begin{eqnarray}
\begin{split}
\left(K^{(0)}_{2N}\right)_{12}=i\left(\bar{\mathbf{u}}[2N]-\varepsilon\bar{\hat{\mathbf{u}}}[2N]\right),
\\
\left(K^{(0)}_{2N}\right)_{21}=i\left( \hat{\mathbf{u}}[2N]-\varepsilon\mathbf{u}[2N] \right).
\end{split}
\label{KN012}
\end{eqnarray}
The relation (\ref{Drel}) yields
\begin{eqnarray}
\det K_{2N}(\lambda)=\det K_{0}(\lambda),
\label{detk}
\end{eqnarray}
where we have used the fact that $d(\lambda)\equiv\det D[2N](x,t,\lambda)$ is independent of $x$ and $t$, and used the equality that $d(\lambda)=d(-\lambda)$. Computing the determinants in the equality (\ref{detk}), we find
\begin{eqnarray}
\begin{split}
\left(K^{(0)}_{2N}\right)_{11}-\varepsilon \left(K^{(0)}_{2N}\right)_{22}=0,
\\
\left(K^{(0)}_{2N}\right)_{11}\left(K^{(0)}_{2N}\right)_{22}
-\left(K^{(0)}_{2N}\right)_{12}\left(K^{(0)}_{2N}\right)_{21}=-\varepsilon\beta^2.
\end{split}
\label{detk1}
\end{eqnarray}
From (\ref{KN012}) and (\ref{detk1}), we obtain
\begin{eqnarray}
\begin{split}
\left(K^{(0)}_{2N}\right)_{22}=\pm i\sqrt{\beta^2 - |\mathbf{u}[2N]-\varepsilon\hat{\mathbf{u}}[2N]|^2},
~~
\left(K^{(0)}_{2N}\right)_{11}=\varepsilon \left(K^{(0)}_{2N}\right)_{22}.
\end{split}
\label{KN01122}
\end{eqnarray}
We need to determine the sign in $\left(K^{(0)}_{2N}\right)_{22}$. From (\ref{kn0}), we obtain
\begin{eqnarray}
\begin{split}
\left(K^{(0)}_{2N}\right)_{22}=\frac{i|\beta|\left(\left|\left(D[2N](0,-t,-\lambda_0)\right)_{11}\right|^2-\left|\left(D[2N](0,-t,-\lambda_0)\right)_{21}\right|^2\right)}
{\left|\left(D[2N](0,-t,-\lambda_0)\right)_{11}\right|^2+\left|\left(D[2N](0,-t,-\lambda_0)\right)_{21}\right|^2}.
\end{split}
\label{KN022}
\end{eqnarray}
The above relation implies that $\left(K^{(0)}_{2N}\right)_{22}$ goes to $i|\beta|$ as $t$ goes to $\infty$, since the DT matrix $D[2N](x,t,\lambda)$ becomes diagonal as $t$ goes to $\infty$ (see e.g. \cite{Gu}). Thus the sign in $\left(K^{(0)}_{2N}\right)_{22}$ should be positive. This completes the proof. \QEDB

For the NLS equation with the zero seed solution, the solution of the associated Lax system at $\lambda=\lambda_j$ takes the form
\begin{eqnarray}
\psi_j(x,t,\lambda_j)\equiv\left(\mu_j,\nu_j\right)^T=e^{-\frac{i}{2}(\lambda_jx-\lambda^2_jt)\sigma_3}\left(u_j,v_j\right)^T, ~~1\leq j\leq N,
\end{eqnarray}
where $\left(u_j,v_j\right)^T$ is a nonzero constant vector. It is easy to check that the following solution of the undressed Lax system corresponding to the parameter $\check{\lambda}_j$,
\begin{eqnarray}
\check{\psi}_j(x,t,\check{\lambda}_j)\equiv\left(\check{\mu}_j,\check{\nu}_j\right)^T=K_0(-\lambda_j)\psi_j(x,t,-\lambda_j),
~~K_0(\lambda)=\diag\left(\lambda-i|\beta|,\lambda+i|\beta|\right),
\label{psirel11}
\end{eqnarray}
satisfies the relation (\ref{psirel2}).
Using the above paired solutions $\left\{\psi_j(\lambda_j),\check{\psi}_j(\check{\lambda}_j)\right\}$, $1\leq j\leq N$, it is straightforward to apply proposition \ref{solution2} to compute soliton solutions of the NLS equation meeting the boundary condition (\ref{tbcnls}). For example, the solution constructed from $\left\{\psi_1(\lambda_1),\check{\psi}_1(\check{\lambda}_1)\right\}$ is
\begin{eqnarray}
\check{u}[1]\equiv u[2]=i\frac{\Delta_1}{\Delta},
\label{solitonnls}
\end{eqnarray}
where
\begin{eqnarray}
\Delta=\left| \begin{array}{cccc}
\lambda_1 \mu_1 & -\lambda_1 \check{\mu}_1 & -\bar{\lambda}_1 \bar{\nu}_1 & \bar{\lambda}_1 \bar{\check{\nu}}_1
\\
\mu_1 & \check{\mu}_1 & -\bar{\nu}_1 & -\bar{\check{\nu}}_1
\\
\lambda_1 \nu_1 & -\lambda_1 \check{\nu}_1 & \bar{\lambda}_1 \bar{\mu}_1 & -\bar{\lambda}_1 \bar{\check{\mu}}_1
\\
\nu_1 & \check{\nu}_1 & \bar{\mu}_1 & \bar{\check{\mu}}_1
\end{array} \right|,
~~
\Delta_1=\left| \begin{array}{cccc}
\lambda^2_1 \nu_1 & \lambda^2_1 \check{\nu}_1 & \bar{\lambda}^2_1 \bar{\mu}_1 & \bar{\lambda}^2_1 \bar{\check{\mu}}_1
\\
\mu_1 & \check{\mu}_1 & -\bar{\nu}_1 & -\bar{\check{\nu}}_1
\\
\lambda_1 \nu_1 & -\lambda_1 \check{\nu}_1 & \bar{\lambda}_1 \bar{\mu}_1 & -\bar{\lambda}_1 \bar{\check{\mu}}_1
\\
\nu_1 & \check{\nu}_1 & \bar{\mu}_1 & \bar{\check{\mu}}_1
\end{array} \right|.
\label{delta1}
\end{eqnarray}
We note that (\ref{solitonnls}) represents single-soliton solution for the boundary problem posed on the positive $x$-axis, despite paired eigenfunctions are used.
This can be understood as follows. The paired eigenfunctions $\left\{\psi_1(\lambda_1),\check{\psi}_1(\check{\lambda}_1)\right\}$ generate two solitons with the same amplitude but opposite velocities (indeed, expression (\ref{u1nls}) implies that the velocity and amplitude of the soliton generated from $\psi_1(\lambda_1)$ are described by the real and imaginary parts of the complex parameter $\lambda_1$, respectively).  Only one of the two solitons is visible on the positive $x$-axis. The boundary conditions can be interpreted as the interaction of the two solitons.

For the complex mKdV equation with the zero seed solution, the solution of the undressed Lax system at $\lambda=\lambda_j$ reads
\begin{eqnarray}
\psi_j(x,t,\lambda_j)\equiv\left(\mu_j,\nu_j\right)^T=e^{-\frac{i}{2}(\lambda_jx-\lambda^3_jt)\sigma_3}\left(u_j,v_j\right)^T, ~~1\leq j\leq N,
\label{psimkdv}
\end{eqnarray}
where, as before, $\left(u_j,v_j\right)^T$ is a nonzero constant vector.
The paired solution of (\ref{psimkdv}) satisfying the relation (\ref{psirel}) is
\begin{eqnarray}
\check{\psi}_j(x,t,\check{\lambda}_j)\equiv\left(\check{\mu}_j,\check{\nu}_j\right)^T=K_0(-\lambda_j)\psi_j(x,t,-\lambda_j),
~~K_0(\lambda)=\diag\left(\varepsilon\left(-\lambda+i|\beta|\right),\lambda+i|\beta|\right).
\label{psirel11}
\end{eqnarray}
With the data (\ref{psimkdv}) and (\ref{psirel11}), it is straightforward to apply proposition \ref{solution} to compute soliton solutions for the complex mKdV equation meeting the boundary conditions (\ref{bccmkdv}).
For example, the solution constructed from $\left\{\psi_1(\lambda_1),\check{\psi}_1(\check{\lambda}_1)\right\}$ reads
\begin{eqnarray}
u[2]=i\frac{\Delta_1}{\Delta},
\label{solitoncmkdv}
\end{eqnarray}
where $\Delta$ and $\Delta_1$ are defined in the same form as (\ref{delta1}) with $\mu_1$, $\nu_1$, $\check{\mu}_1$ and $\check{\nu}_1$ replaced by (\ref{psimkdv}) and (\ref{psirel11}).
It is worth mentioning that the solution (\ref{solitoncmkdv}) stands for $2$-soliton solution, since the paired eigenfunctions $\left\{\psi_1(\lambda_1),\check{\psi}_1(\check{\lambda}_1)\right\}$ in this case create two solitons with the same velocity and amplitude (the initial position is different). In other words, starting from the zero seed solution, only even numbers of solitons meeting the boundary conditions (\ref{bccmkdv}) can be constructed via this manner. This is the essential difference with respect to the case of the NLS equation with the boundary condition (\ref{tbcnls}). We also note that it is possible to construct odd numbers of solitons meeting the boundary conditions (\ref{bccmkdv}) by using our dressing method, once one find a single-soliton meeting the boundary conditions (\ref{bccmkdv}) as a seed solution. However, the main challenge in doing so is to solve the corresponding Lax system to obtain the paired eigenfunctions as required in proposition 6. We will not discuss this issue in the present paper.

\section*{ACKNOWLEDGMENTS}
This work was supported by the National Natural Science Foundation of China (Grant No. 12271221).

\vspace{1cm}
\small{

}
\end{document}